\documentclass{vgtc}                          

\ifpdf
  \pdfoutput=1\relax                   
  \usepackage{graphicx}                
  \DeclareGraphicsExtensions{.pdf,.png,.jpg,.jpeg} 
\else
  \usepackage{graphicx}                
  \DeclareGraphicsExtensions{.eps}     
\fi%

\graphicspath{{figures/}{pictures/}{images/}{./}} 

\usepackage{microtype}                 
\PassOptionsToPackage{warn}{textcomp}  
\usepackage{textcomp}                  
\usepackage{mathptmx}                  
\usepackage{times}                     
\usepackage{cite}                      
\usepackage{tabu}                      
\usepackage{booktabs}                  

\usepackage{amssymb}
\usepackage[caption=false]{subfig}
\usepackage{caption}
\usepackage{lipsum}
\usepackage{algorithmicx}
\usepackage[noend]{algpseudocode}
\usepackage{algorithm}
\usepackage{stfloats}
\usepackage{float}
\usepackage{gensymb}
\usepackage{color}

\onlineid{1186}

\vgtccategory{Research}

\vgtcinsertpkg

\title{Overlap-free Drawing of Generalized Pythagoras Trees\\ for Hierarchy Visualization}

\author{\parbox{13em}{\centering Tanja Munz\thanks{e-mail: tanja.munz@visus.uni-stuttgart.de}\\ 
    \scriptsize VISUS, University of Stuttgart}
\and \parbox{13em}{\centering Michael Burch\thanks{e-mail: m.burch@tue.nl}\\
    \scriptsize Eindhoven University of Technology}
\and \parbox{13em}{\centering Toon van Benthem \\
    \scriptsize Eindhoven University of Technology}
\vspace{0.5em}
\and\parbox{13em}{\centering Yoeri Poels \\
    \scriptsize Eindhoven University of Technology}
\and \parbox{13em}{\centering Fabian Beck \thanks{e-mail: fabian.beck@paluno.uni-due.de}\\ 
    \scriptsize University of Duisburg-Essen}
\and \parbox{13em}{\centering Daniel Weiskopf \thanks{e-mail: daniel.weiskopf@visus.uni-stuttgart.de}\\ 
    \scriptsize VISUS, University of Stuttgart}
}

\teaser{
  \centering
  \includegraphics[width=0.99\textwidth]{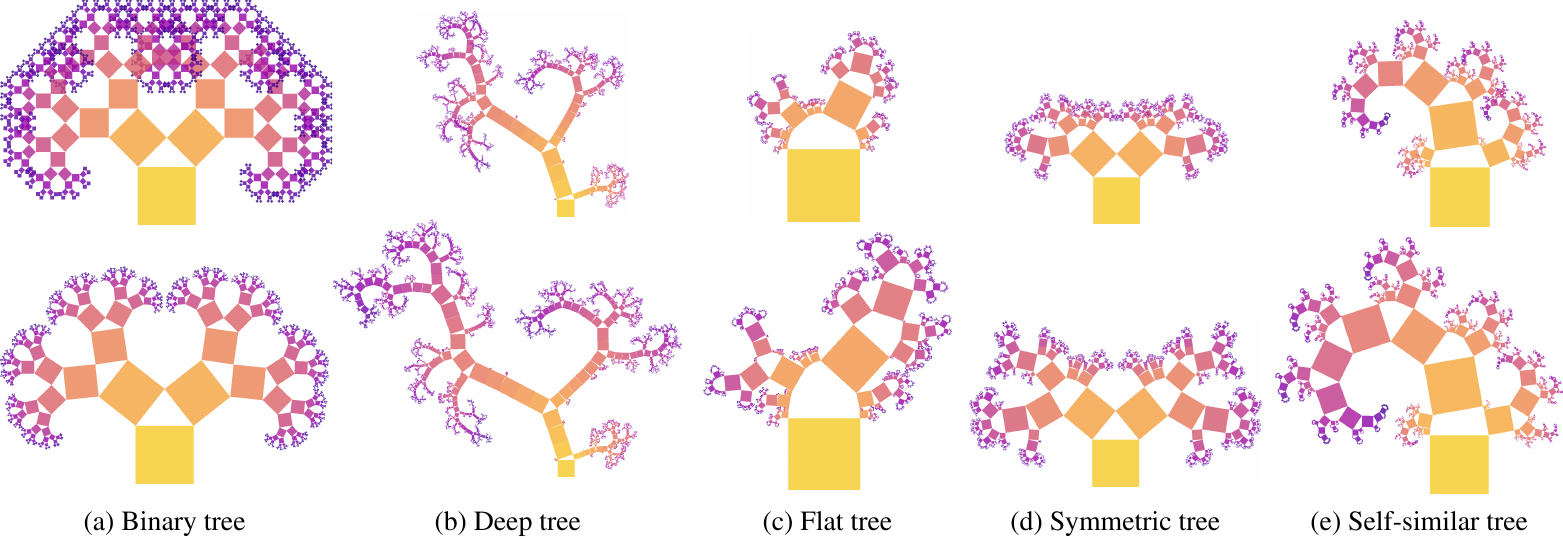}
  \caption{Comparing the original drawing algorithm of generalized Pythagoras trees (top) to equivalent ones from the suggested overlap-free approach (bottom). The overlap present in the original drawings is completely removed.}
  \label{fig:teaser}
}

\abstract{Generalized Pythagoras trees were developed for visualizing hierarchical data, producing organic, fractal-like representations. However, the drawback of the original layout algorithm is visual overlap of tree branches.
To avoid such overlap, we introduce an adapted drawing algorithm using ellipses instead of circles to recursively place tree nodes representing the subhierarchies. Our technique is demonstrated by resolving overlap in diverse real-world and generated datasets, while comparing the results to the original approach.
} 

\CCScatlist{
  \CCScatTwelve{Human-centered computing}{Visualization}{Visualization application domains}{Information visualization}
}

\begin{document}


\firstsection{Introduction}

\maketitle

\begin{figure}[b]
    \centering
    \includegraphics[width=0.47\textwidth]{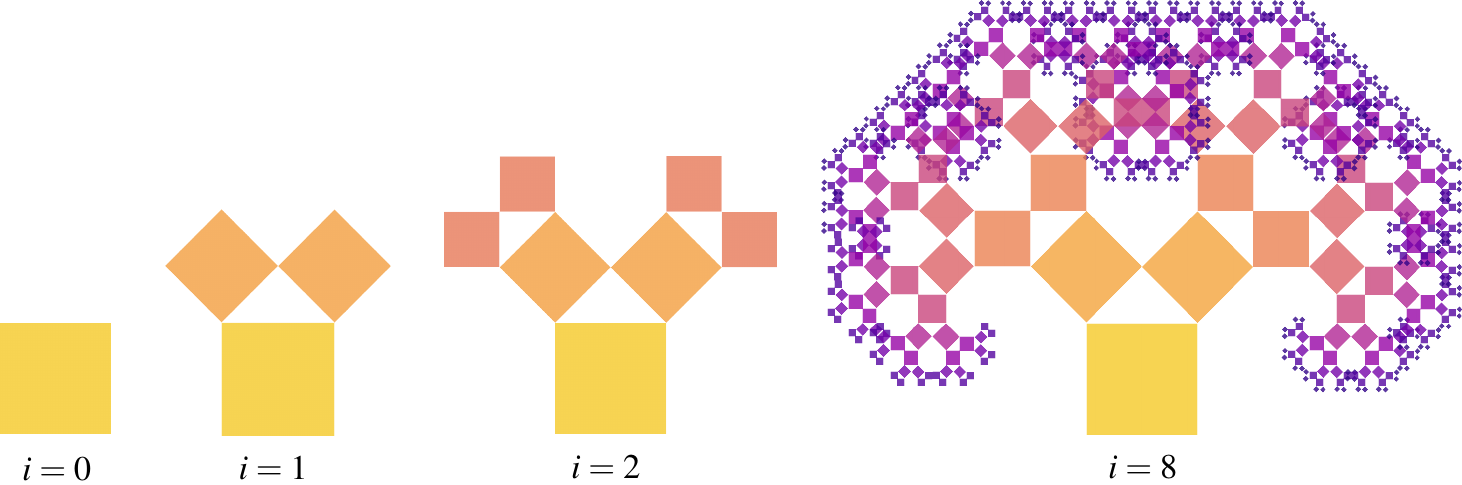}
    \caption{Iterative construction of a Pythagoras tree fractal.}
    \label{fig:pyth_fractal}
\end{figure}

The Pythagoras tree~\cite{Bosman:57} is a fractal constructed from squares. On top of a square, two smaller squares are placed with their lower corners on a semicircle so that their bottom sides together with the top side of the bigger square form a Pythagorean triangle. \autoref{fig:pyth_fractal} illustrates this recursive process.

Generalized Pythagoras trees~\cite{pythagorastree, pythagorastree2} are a modification thereof to visualize hierarchical data. In the iterative process, branches are left out if respective subhierarchies do not exist in the data and non-binary branches provide the option to visualize general hierarchies. The resulting visualizations form organic shapes and show self-similarities~\cite{pythagorastree}. However, the visualization technique has the drawback that occasional overlap of branches reduces its readability (see \autoref{fig:teaser}, top).

To solve the overlap issue, this paper introduces a novel extension: \emph{overlap-free} generalized Pythagoras trees (see \autoref{fig:teaser}, bottom). This adopts ideas from force-directed graph layout~\cite{forcedirected} and applies forces to stretch or compress the semicircle on which the smaller squares are positioned (i.e., the semicircle becomes a semi-ellipse). In this way, whenever an overlap is detected, the respective branches are pushed away from each other and become narrower.
Simulating the forces and movements over a number of iterations, the layout reaches a stable, overlap-free state (see \autoref{fig:tax}). We argue that this substantially improves the readability of generalized Pythagoras tree visualizations while preserving their visual properties. 

\section{Related Work}
\label{sec:related}

Many different visualization techniques are available to represent hierarchies. These techniques use a variety of approaches to arrange and connect parent nodes with their children. Common methods use the concepts of node-link diagrams, stacking, indentation, or nesting (e.g., \cite{kruskal1983icicle, burch2010indented, shneiderman1992tree}). Schulz~\cite{Schulz2011} provides a survey.

As mentioned above, our work is based on a specific hierarchy visualization technique: the generalized Pythagoras tree \cite{pythagorastree, pythagorastree2}.
CactusTrees \cite{cactusTree} use a similar base approach to generalized Pythagoras trees for placing vertices represented by circles. A scale factor between parent nodes and children is used to handle overlapping. This, however, decreases the size of subhierarchies, which also reduces readability. While it is possible to remove all collisions, it is questionable how well detailed information can still be retrieved from areas previously occupied by overlap, which are then rendered much smaller.
Other nature-oriented tree structures can be created with Colorful Trees~\cite{Haensch19RFVis}. They also have the problem of overlapping branches and leaves but solve the issue by drawing the hierarchy only up to a user-defined depth and by using pie charts as a representation for subhierarchies.
In our approach, we will neither hide and summarize information nor just decrease the size of nodes to reduce overlaps. Instead, we will change the size of vertices only when necessary to reduce clutter.

For node-link diagrams, graph layout algorithms are popular to address visual clutter. Most notable are force-directed graph layout algorithms (e.g., by Fruchterman and Reingold \cite{forcedirected} or Kamada and Kawai \cite{kamadaKawai}), which apply attraction and repulsion forces to get a readable layout. The technique we propose in this paper uses a similar approach that is based on different forces to reduce overlaps while preserving the tree's general layout.
Gansner and Hu~\cite{GansnerNodeOverlaop} investigated, in a similar iterative approach, node overlap removal for graphs with different node sizes. Their goal is also to maintain the original structure.

Related approaches can also be found in application-specific work. Algorithms that deal with overlap removal are required for Ribonucleic Acid (RNA) structures \cite{WiegreffeRNA, AuberRNA}. These structures are represented by RNA secondary structures that can be drawn as planar graphs with some constraints. For example, RNApuzzler \cite{WiegreffeRNA} uses a similar algorithmic solution to remove overlap by relaxing some constraints; the algorithm recursively checks and solves ancestor and sibling intersections.

\section{Visualization Technique}
\label{sec:vis}

Let $G=(V,E)$ denote the tree (hierarchy) to be visualized, where $V$ is the set of nodes and $E \subseteq V \times V$ is the set of directed edges that define the tree.
We allow positive weights to be augmented to the vertices. This is modeled as a function $w: V \rightarrow \mathbb{R^+}$. The weight of a vertex does not necessarily need to be the sum of its children.

\subsection{Modifying Tree Shape}
\label{sec:ellipse}

\begin{figure}
    \centering
    \subfloat[All children are assigned 45\degree, resulting in different sizes.
    \label{fig:ellipse_preadjust}]{\includegraphics[width=0.394\columnwidth]{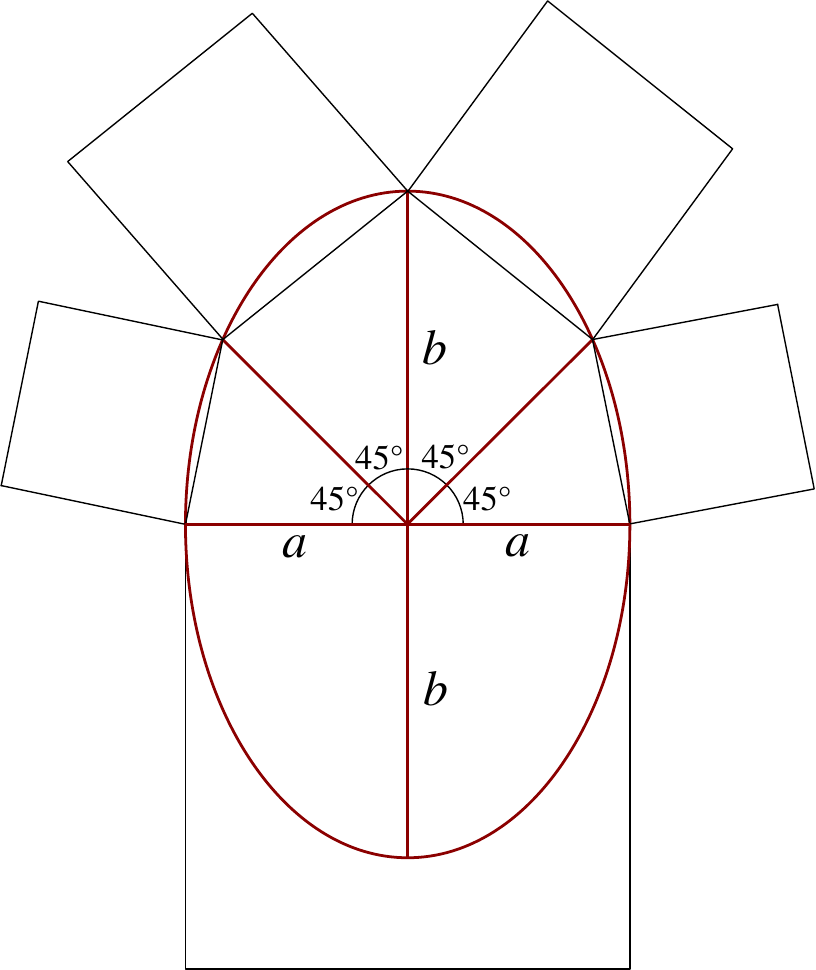}}\qquad
    \subfloat[Angles are rescaled to make children the same width.
    \label{fig:ellipse_postadjust}]{\includegraphics[width=0.416\columnwidth]{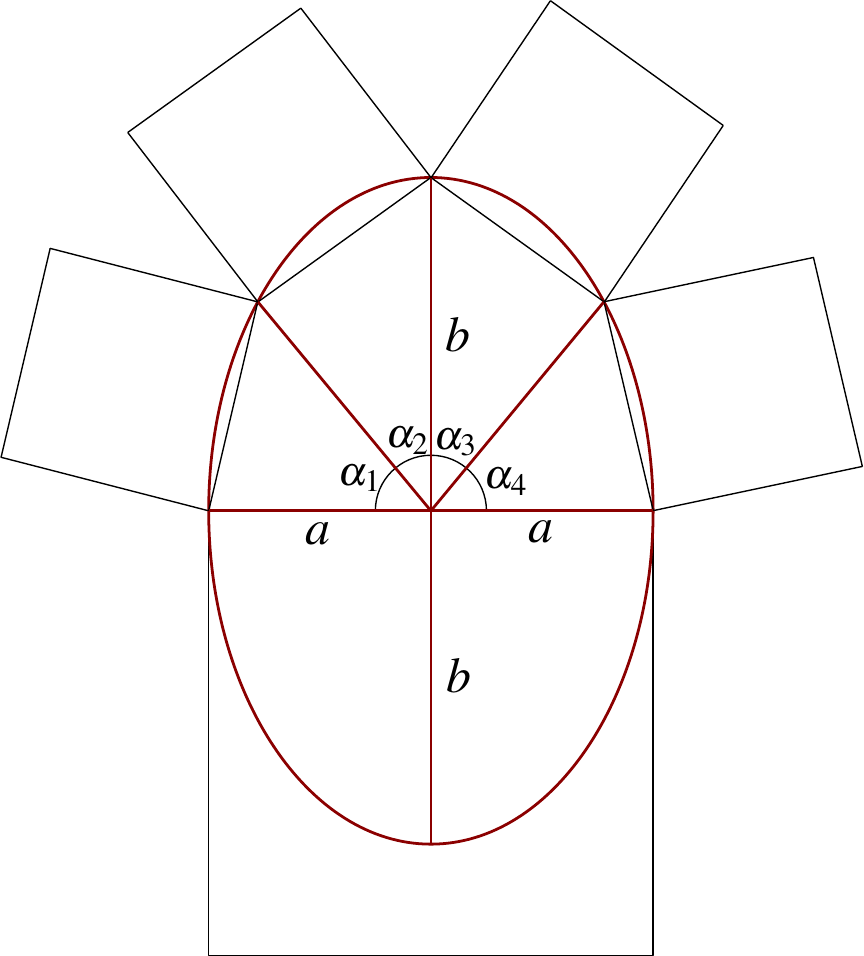}}\qquad
    \caption{Finding children positions using an ellipse circumference for four children with equal weights.}
    \label{fig:ellipse}
\end{figure}

To avoid overlap we propose a novel method for modifying the tree structure. Rather than positioning the children of a node according to the circumference of a semicircle around the top of a rectangle, we position them based on the circumference of a semi-ellipse. This ellipse is defined as
$x^2/a^2 + y^2/b^2 = 1$
with $a$ and $b$ being the two semi-axes perpendicular to each other, of the ellipse (see \autoref{fig:ellipse}).
By using an ellipse instead of a circle, we keep the tree connected and maintain its round shape. Additionally, notice that an ellipse with parameters $a=1, b=1$ is a circle.

We use an iterative process to position children on the ellipse as we should not simply divide the circumference of the semi-ellipse by dividing 180\degree ~over the weights as it is possible for a semicircle (\autoref{fig:ellipse_preadjust}).
We compare the assigned width of a child to the weight it was assigned in relation to the sum of widths of all children, and iteratively re-scale the angles (\autoref{fig:ellipse_postadjust}).

\begin{algorithm}[t]
\caption{Force-directed Generalized Pythagoras Tree}\label{alg:thealg}
\begin{algorithmic}[1]
\Function{FDGPT}{$G = (V, E)$}
    \State Initialize a window query structure $Q$
    \State $C\gets\bigcup\limits_{v\in V}Q.$query$(v)$
    \While {$C$ \textbf{not} empty}
        \For {$(u, v) \in C$}
            \State Find first common ancestor $z$ of $u$ and $v$
            \State $z.$spread $\gets z.$spread + 1
            \For{all nodes $z$ on paths from $v, u$ to $z$}
                \State $z.$narrow $\gets z.$narrow + 1
            \EndFor
        \EndFor
        \For{all nodes $v\in V$}
            \If {$v.$spread $> v.$narrow}
                \State $v.b \gets$ min$(1.1 \cdot v.b, \phi)$ \Comment{Push Force}
            \ElsIf {$v.$spread $< v.$narrow}
                \State $v.b \gets 0.9 \cdot v.b$ \Comment{Pull Force}
            \EndIf
            \State $v.b \gets v.b + (1 - v.b) \cdot v.lr$ \Comment{Neutral Force}
            \State $v.lr \gets v.lr\cdot 0.9$
            \State Q.update(v)
        \EndFor
        
        \State $C\gets\bigcup\limits_{v\in V}Q.$query$(v)$
    \EndWhile
\EndFunction
\end{algorithmic}
\end{algorithm}

\begin{figure*}[h]
    \centering
      \subfloat[Overlapping structure.
      \label{fig:tree1}]{\includegraphics[width=0.2217\textwidth]{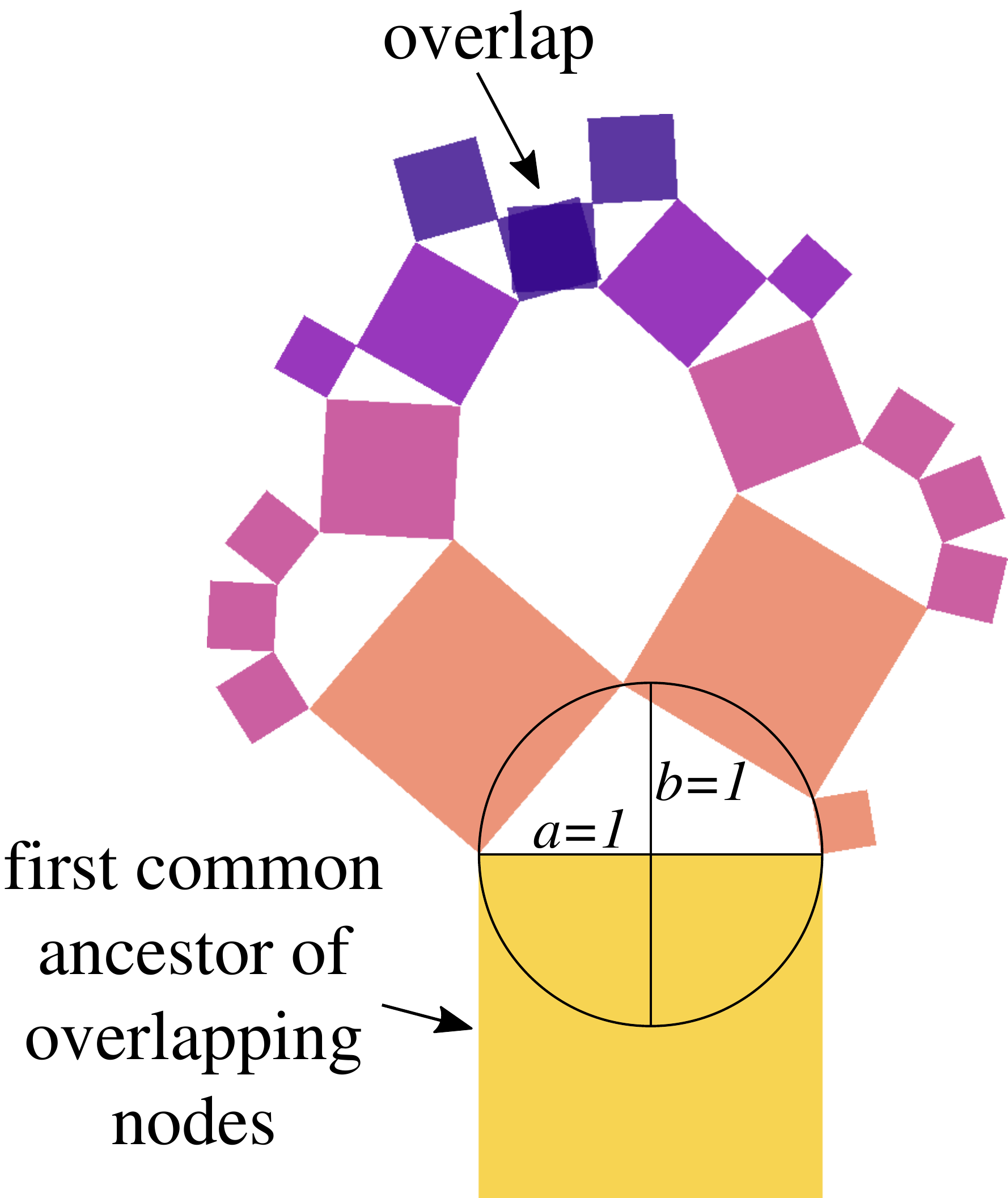}}\qquad
      \subfloat[The result of increasing $b$ in the first common ancestor.
      \label{fig:tree2}]{\includegraphics[width=0.2293\textwidth]{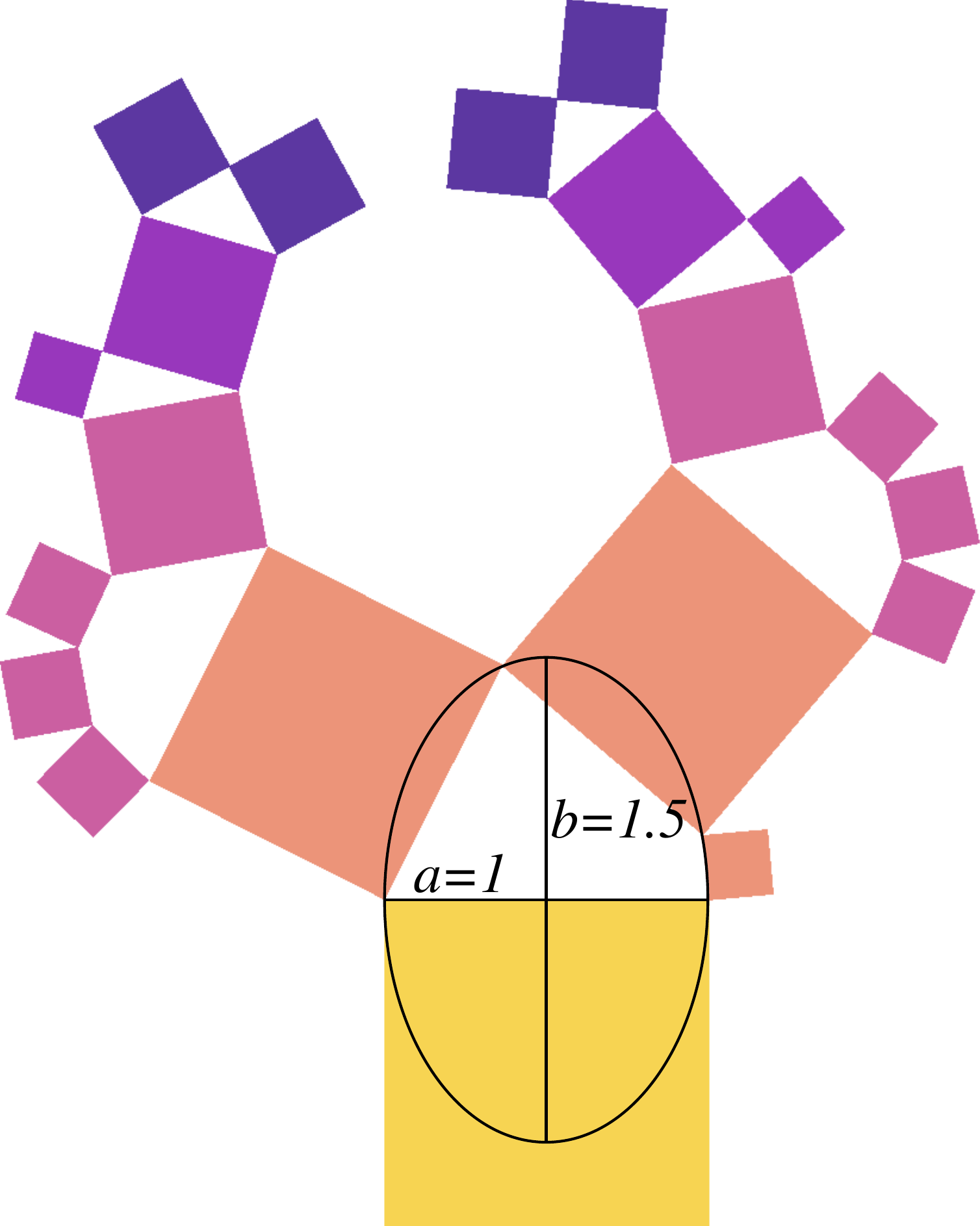}}\qquad
      \subfloat[The result of decreasing $b$ in one of the subtrees.
      \label{fig:tree3}]{\includegraphics[width=0.185\textwidth]{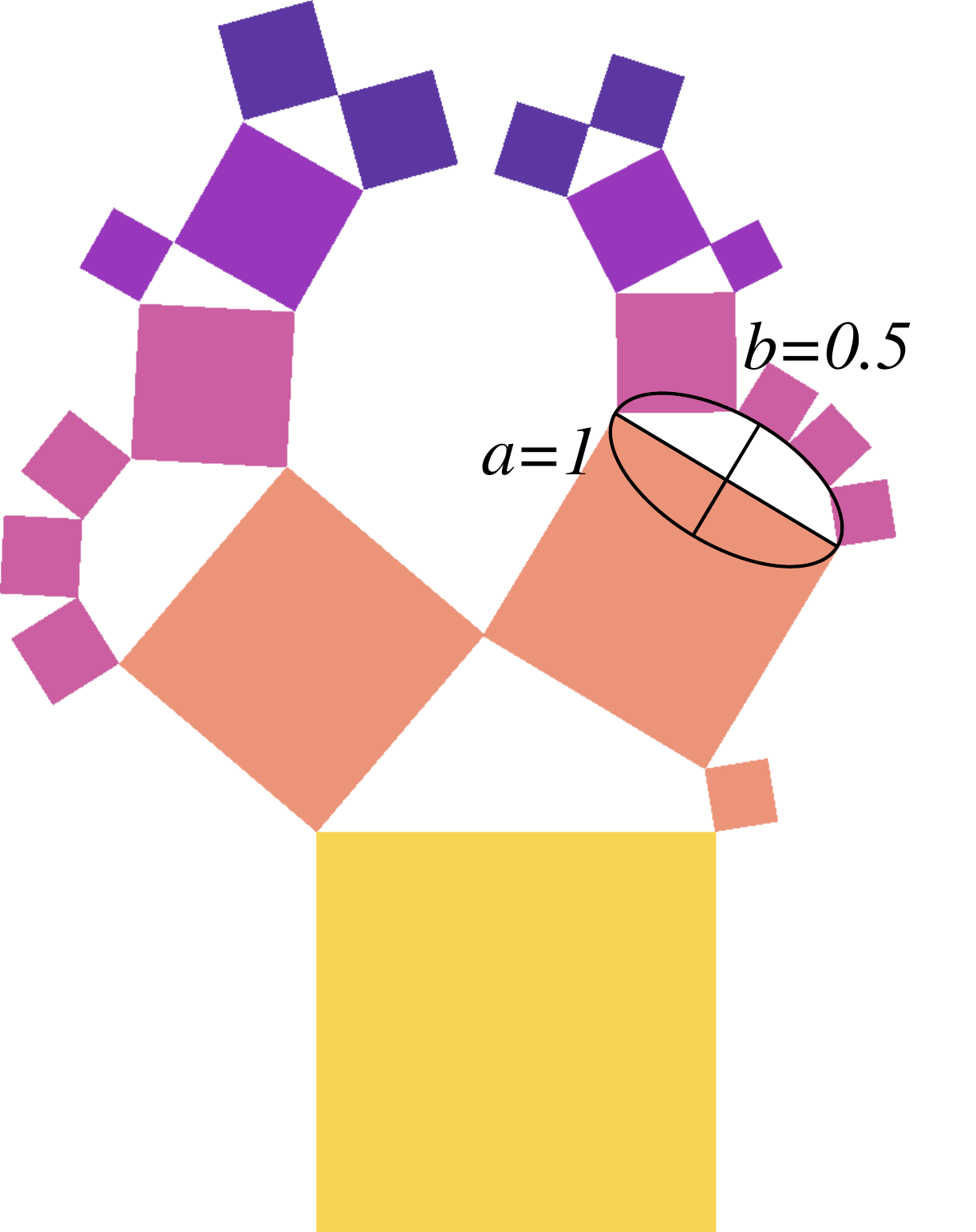}}
     \caption{Two ways of reducing overlap.}
     \label{fig:overlapexp}
\end{figure*}

The usage of ellipses allows us to modify $b$ for a given node. We illustrate this in \autoref{fig:tree1}. By increasing $b$, the ellipse becomes taller, and subtrees are pushed apart, which also results in larger subtrees (\autoref{fig:tree2}). Decreasing $b$ reduces the height of an ellipse, resulting in a narrower subtree, as child trees are pulled together. Additionally, the size of subnodes decreases.
If $b = 0$ for all ellipses, the structure transforms into an icicle plot \cite{kruskal1983icicle}.

\subsection{Finding Overlap}

To efficiently find where the tree overlaps in the original layout, we store the nodes in a quadtree $\mathcal{T}$. For each rectangle, we perform a window query based on its axis-parallel bounding box on $\mathcal{T}$ to find candidate overlapping rectangles. Finally, we confirm for all candidate rectangles whether they actually overlap.
The quadtree structure improves iteration time. Compared to a naive overlap check taking $O(n)$ per node, we now take $O(d+k)$, where $d$ is the quadtree depth, and $k$ is the number of points returned~\cite{berg}.

\subsection{Force Model}
\label{sec:forcemodel}

To remove overlap, we use a hit-based approach, and look at the issue from overlapping nodes $u, v$. The general idea is that we want to spread out the two subtrees that contain $u$ and $v$, while also making each of them narrower.

To do this, we find $z$, the first common ancestor of $u$ and $v$, and increment a \textit{spread} counter on it. Then, we look at the nodes on the paths from $u$ and $v$ to $z$, incrementing a \textit{narrow} counter for each.
Now, we apply three types of forces: For each node, we check whether it is marked more often to spread or to narrow. If spread dominates, we apply a \emph{push force} that increases the $b$ parameter of the ellipse by 10\%, up to a maximum of $\phi$, the golden ratio (1.618)~\cite{phi}.
If narrow dominates, we apply a \emph{pull force} that decreases $b$ by a factor of 10\%.
Finally, we apply a neutral force that directs $b$ back to $1$, aiming to reduce the number of extreme values in our tree. This force becomes smaller as more iterations modify the nodes, indicated by the learning rate parameter ($lr$) that is initialized with $0.1$. Initially, this neutral force is a limiting factor to how flat or pointy an ellipse can get, but after it decreases sufficiently $b$ can decrease as far as nearly 0, resulting in a flat stack of subtrees. This occurs only in extreme cases, as shown in \autoref{sec:case}.

Pseudocode for the stabilization can be found in Algorithm \autoref{alg:thealg} and supplementary material contains a time complexity analysis.

\begin{figure*}[b]
 \centering
      \subfloat[The tree with no modifications: 51,930 collisions.\label{fig:tax1}]{\includegraphics[width=0.23\textwidth]{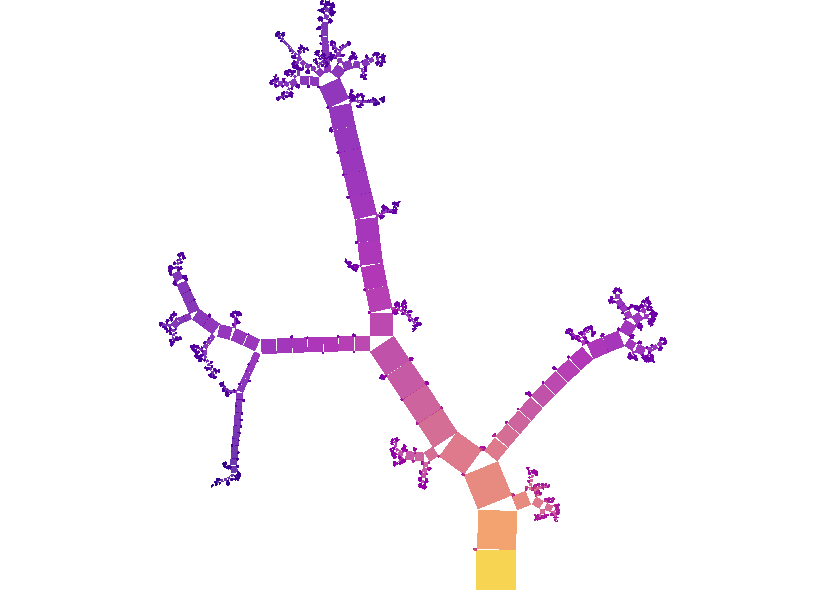}}\quad
      \subfloat[The tree after just 6 iterations: Still 39,192 collisions. \label{fig:tax2}]{\includegraphics[width=0.23\textwidth]{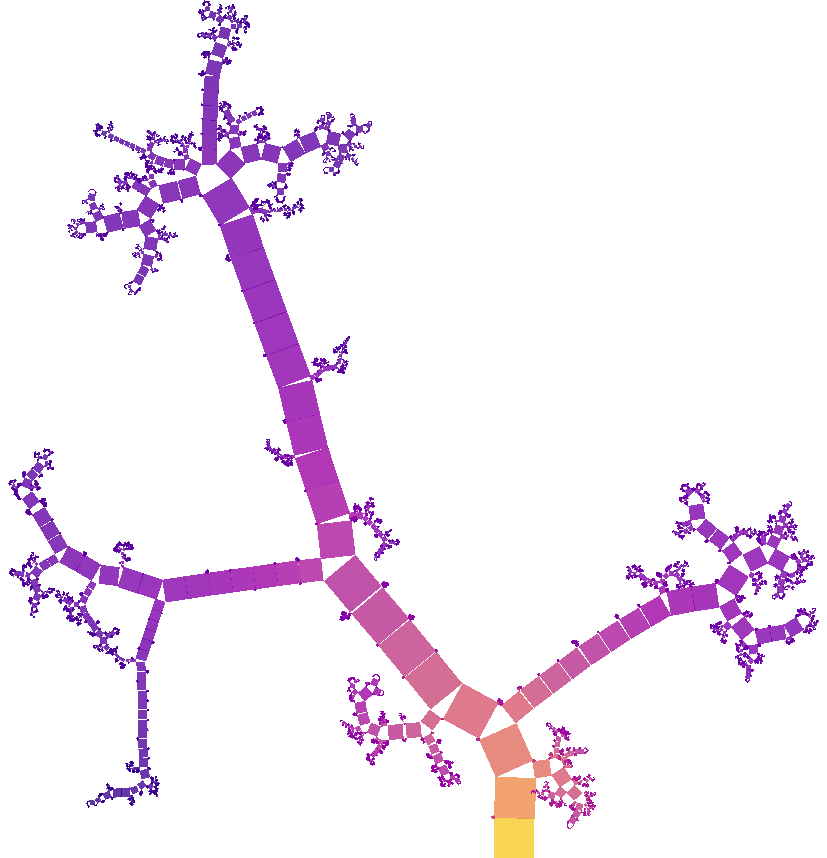}}\quad
      \subfloat[The tree after 155 iterations: All overlap has been resolved. \label{fig:tax3}]{\includegraphics[width=0.23\textwidth]{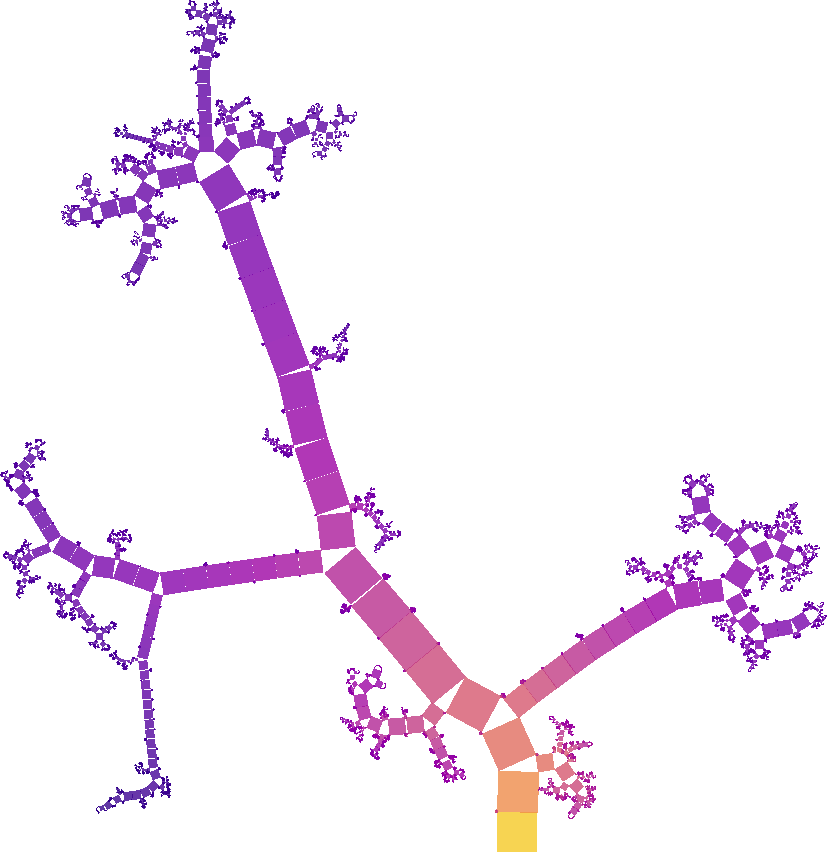}}\quad
      \subfloat[Solved with alternative height settings. \label{fig:taxl}]{\includegraphics[width=0.23\textwidth]{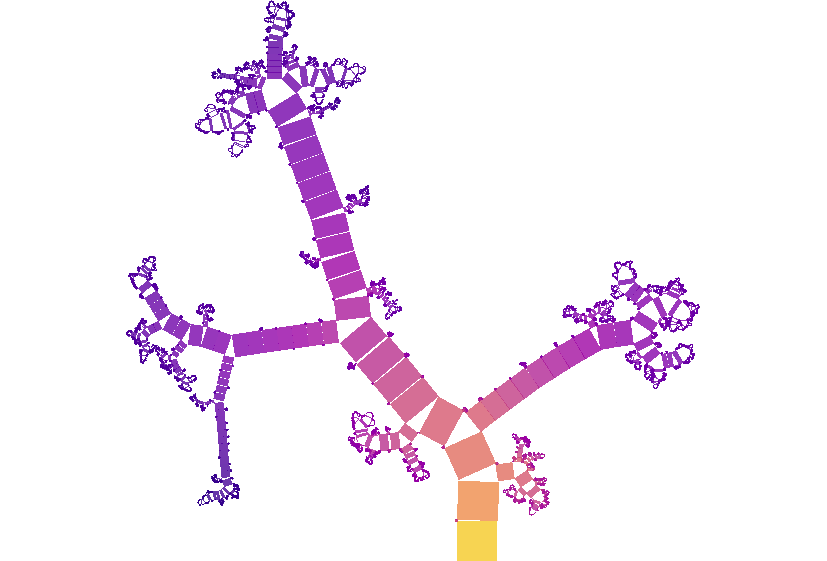}}
      \caption{The NCBI taxonomy dataset as generalized Pythagoras tree: Resolving the overlap (a--c) and the result for different height settings.}
      \label{fig:tax}
\end{figure*}

\section{Examples and Discussion}
\label{sec:case}

To evaluate the outcome of our approach, we apply it to typical tree examples. The overlap-free trees are compared to the initial tree, and we discuss questions of geometry, perception, readability, and aesthetics. To clearly demonstrate differences in scale, root nodes (of the same tree) will be of same size.
In these examples, we do not encode extra information in the height of the nodes, but rather kept them as squares. We will, however, briefly touch upon the influence the height can have in \autoref{sec:taxonomy}. Furthermore, the color of the nodes can also be used to encode specific information. In our case, it is a simple color gradient depicting the depth of the node. Lastly, unless stated otherwise, the weight of the node is assigned according to the number of nodes in its subtree.

\subsection{NCBI Taxonomy Tree}
\label{sec:taxonomy}

The first dataset is the NCBI taxonomy with organism names and taxonomic lineages \cite{Federhen2012}. The generalized Pythagoras tree from this dataset is depicted in \autoref{fig:tax1}. It has a total of 324,269 nodes. Initially, 51,930 collisions between nodes occur. Fully resolving all collisions takes 155 iterations. First, the tree substantially increases in size, however when the limit for the $b$ parameter of the larger nodes is reached it stabilizes and fixes overlap on a smaller scale, as is visible in the (lack of) difference between \autoref{fig:tax2} and \autoref{fig:tax3}. 

The overall shape of the tree is maintained, allowing for an overview of the data with no interaction necessary\textemdash besides zooming and translating the view. The relative scale of nodes is impacted to a certain extent, meaning one must be careful when directly comparing nodes in terms of size.

\begin{figure}[t]
    \centering
    \includegraphics[width=0.4\textwidth]{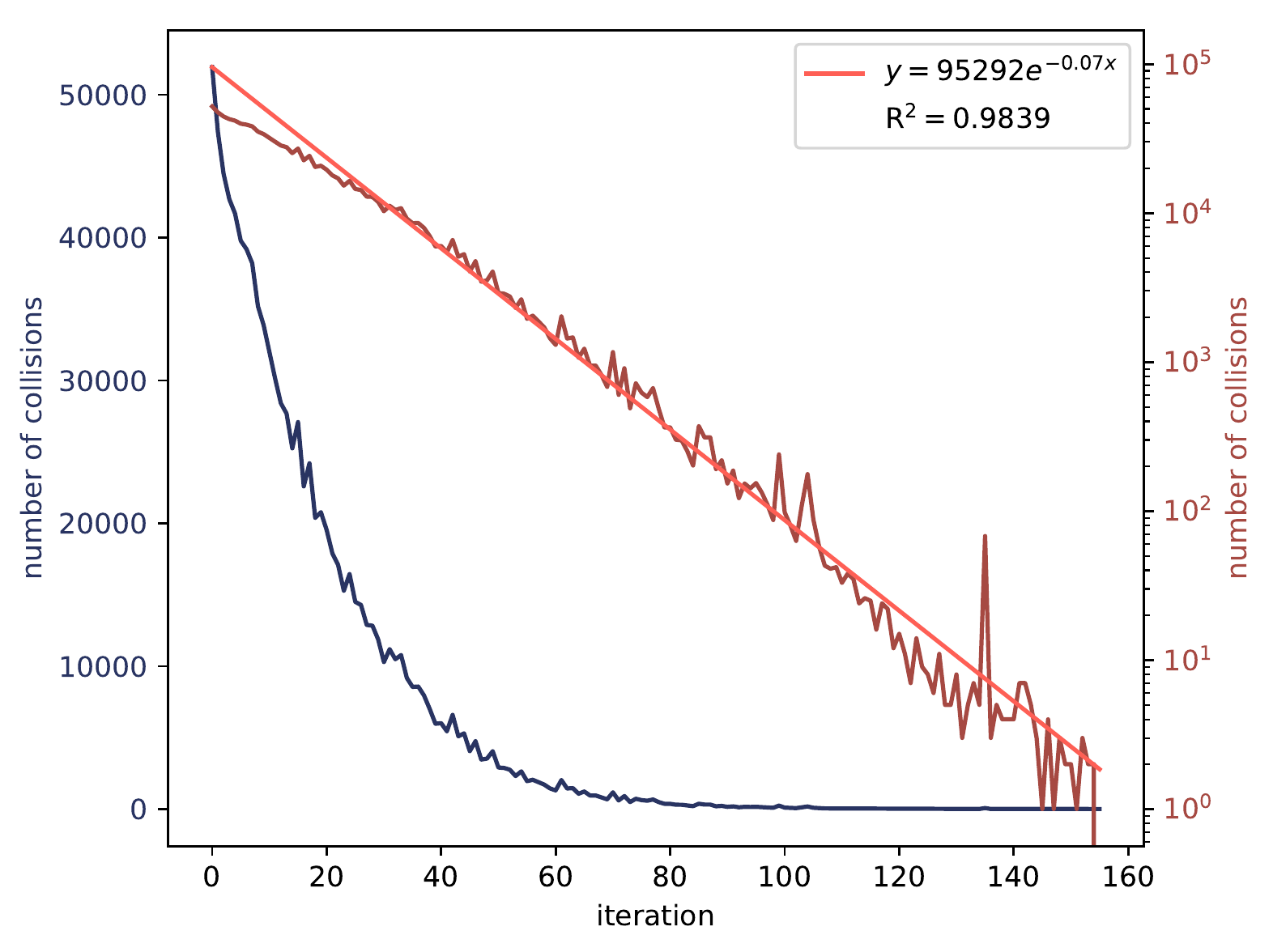}
    \caption{Collision number relative to the iterations of the algorithm.}
    \label{fig:it_combined}
\end{figure}

An overview of the number of collisions relative to the number of iterations is shown in \autoref{fig:it_combined}. It hints at a negative exponential relation between the number of collisions and the iterations, which becomes clear when plotted against a logarithmic scale.

We also experimented with limiting the height of the nodes: Rather than keeping them square-shaped, we set the height to $min(original\_height, width)$. Solving with these parameters led to a similar runtime: 158 iterations for a complete solution, with a similar negative exponential decrease in collisions relative to the iteration (R$^2 = 0.9803$ for a simple exponential line fit). The resulting tree is depicted in \autoref{fig:taxl}. A clear trade-off is visible here: by limiting the height, the tree does not expand nearly as much, using roughly the same space to depict all information. This comes at the cost of abandoning the original structure. Changes in shape on a node-basis means that similarities that might occur in hierarchical shape are harder to recognize; the tree shape is more impacted by the force-directed modifications. Furthermore, one might consider the square-shaped trees more aesthetically pleasing, as the counterpart's nodes can far exceed width--height ratios considered visually appealing\cite{fechner1997square}.

\subsection{File System Hierarchy}

\begin{figure}[t]
    \centering
    \includegraphics[width=0.99\linewidth]{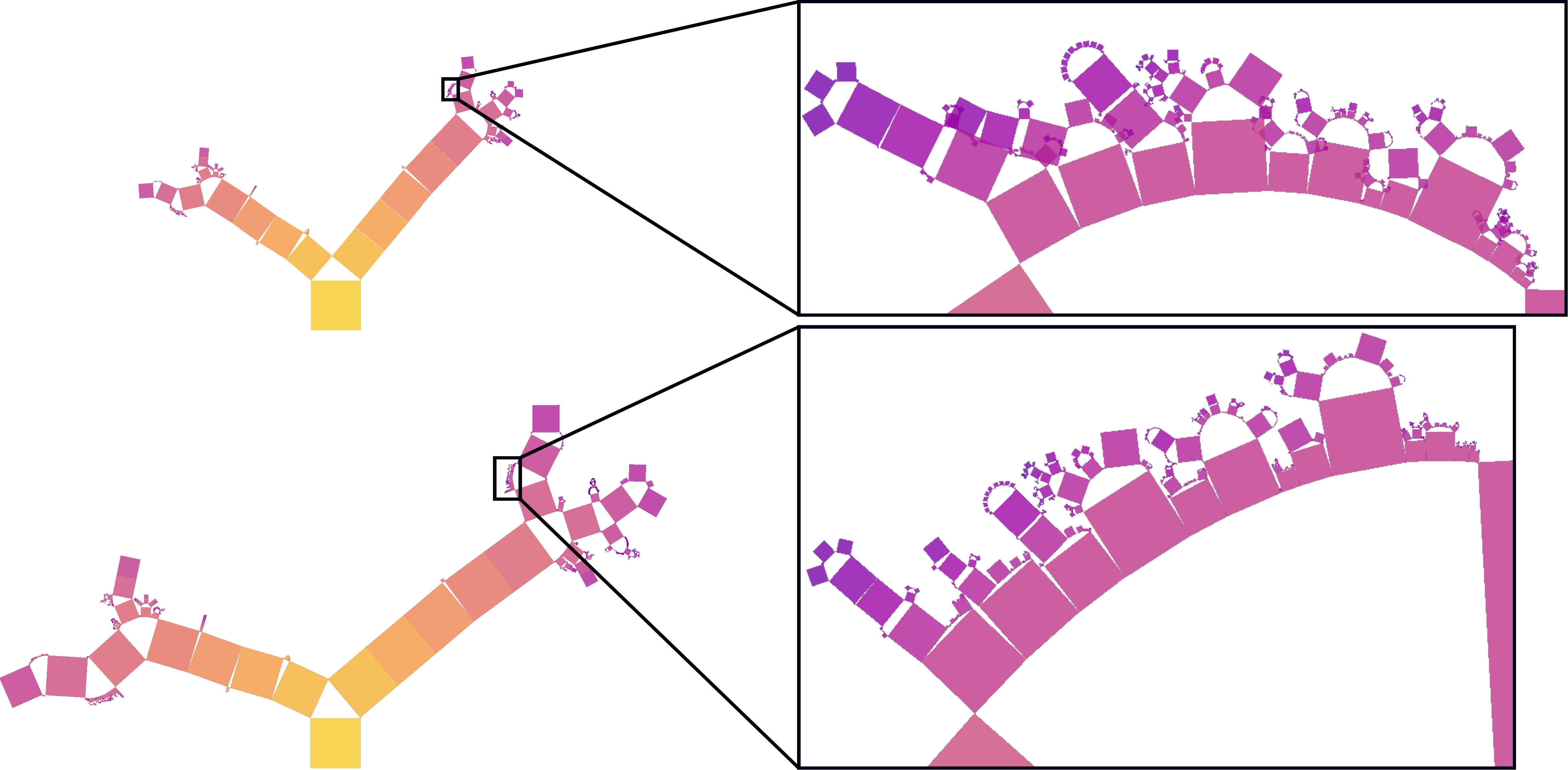}
    \caption{The file system hierarchy (top) and after resolving overlap (bottom). Zoomed-in section shown on the right.}
    \label{fig:filetree}
\end{figure}

A second dataset we consider is a file system hierarchy containing the visualization tool and the paper-related files, depicted as a generalized Pythagoras tree in \autoref{fig:filetree} (top). Nodes are scaled according to file size rather than number of descendants. It has 30,859 nodes in total, with 8,483 collisions in its initial state. It took a total of 164 iterations to fully resolve the overlap, with the result depicted in \autoref{fig:filetree} (bottom). Similar alterations as for the taxonomy tree occur, with the overlap reduction being traded for a slight increase in size. 

A disadvantage of the current method is that, when many nodes are stacked close together, it is difficult to maintain a Pythagoras tree-like structure without overlap. As a result, the $b$ parameter of nodes gets decreased to the point where nodes are almost stacked on top of each other (icicle plot), deviating from the original structure. This is quite apparent in the subtree of the file system hierarchy depicted in \autoref{fig:filetree} on the right side.

\subsection{Artificially Generated Hierarchies}

Lastly, we consider a set of generated hierarchies to discuss the influence on geometry, perception, readability, and aesthetics. An overview of all these trees and their overlap-free counterparts is depicted in \autoref{fig:teaser}.

As the resolved trees are still based around a similar method, i.e., the placing of the children along a circle/ellipse perimeter on their parent node, the visual encoding for displaying relationships between nodes does not change. It still applies that related rectangles being arranged on an imaginary curve get connected by the human reader subconsciously (as described by Beck et al.~\cite{pythagorastree}).

Something that might suffer is the uniqueness of subtrees: Whereas self-similar trees remain recognizable as such, a binary tree quite drastically changes its shape. What might represent the same hierarchical structure could be visualized quite differently after resolving overlap. It is likely that, when comparing non-overlapping trees only, the fingerprint of the tree will still be valuable, as subtrees locally always resolve in the same manner.

The readability benefits from the resolvement of overlap. We can better distinguish hierarchical structures: for the example of the binary tree (\autoref{fig:teaser}a), in the initial state, it is difficult to tell which child belongs to which subtree, as multiple nodes overlap at the same time. Similarly, for the deep hierarchy (\autoref{fig:teaser}b), it is easier to tell the branches apart, as they do not cross each other anymore.

As mentioned previously, the main structure remains the same compared to generalized Pythagoras trees, thus sharing some attributes regarding readability and scalability: Our approach can be used to display large datasets in a readable manner, however, suffers when being used to display flat hierarchies, as nodes decrease in size fast~\cite{pythagorastree}. A disadvantage is that the result becomes slightly larger in almost all cases. This is especially visible in the deep, flat, symmetric, and self-similar hierarchies (\autoref{fig:teaser}b--e). The additional space, however, is quite restricted; hence, it is  unlikely that this becomes a limiting factor.
The fractal-like structure between overlapping and non-overlapping trees is quite similar. From this, we can draw the conclusion that the aesthetic quality of one is likely not too different from the other~\cite{SPEHAR2003813, forsythefractal}.
Additionally, we refer the reader to the \textit{Discussion} section of the original paper \cite{pythagorastree2} for more details on the characteristics of generalized Pythagoras trees.

\subsection{Iterations and Termination}\label{sec:resoverview}

A point of interest is whether a relation between the input state and the number of iterations exists. Unfortunately, for the trees used in this section (and a set of additional artificial trees to increase the sample size), we could not establish a clear relation (see supplementary material).
The behavior of the number of collisions relative to the iterations seems somewhat similar for most solutions, however, all showing some (albeit somewhat erratic) negative exponential relation (see supplementary material). If resolving all overlap is not mandatory, one could choose to run the algorithm for a limited number of iterations, as the number of overlapping nodes initially drops drastically, resolving most overlap for relatively little cost.

For all datasets we tested, the algorithm terminated and created overlap-free structures.
Generally, the relaxation allows the reduction of all $b$ values to zero, resulting in a convergence to an icicle plot, which is overlap-free.
However, there might still be cases with repeated increase and decrease of b. A formal proof for termination remains future work.

\section{Conclusion}
\label{sec:conclusion}

In this paper, we presented an algorithm for the generalized Pythagoras trees to visualize hierarchies without overlap. Drawing nodes on the perimeter of an ellipse allows modification of the shape of (sub)trees. Using three force-based rules, we iteratively change the shape of the tree until no collisions are found. The algorithm is able to produce visualizations without overlap on all input sets that were tested. It does so without compromising on readability and aesthetics, leading us to believe that this approach is superior to the original layout.

\acknowledgments{
This work was funded by the Deutsche Forschungsgemeinschaft (DFG, German Research Foundation) under Germany's Excellence Strategy -- EXE-2075 -- 390740016.}

\bibliographystyle{abbrv-doi}

\bibliography{ms}
\end{document}


\maketitle

\subsection*{Algorithm Analysis}
In the following, we will analyze the time complexity of Algorithm 1 as described in Section 3.3 of the main document. We found no explicit relation between the input state and the number of iterations until stabilization, as shown by the empirical evidence (see plots in next section). Therefore, we will analyze the complexity of a single iteration. That is: lines 5-18 of Algorithm 1.

We divide the analysis into the three subroutines of the algorithm: (1) finding overlapping nodes; (2) finding common ancestors and increasing `spread'/`narrow'-counters; (3) applying the forces.

For each of $n$ nodes in our tree we perform a window query on our quadtree structure. This query takes $O(d+k)$, where $d$ is the depth of the quadtree and $k$ is the number of points returned. Thus, for each iteration finding all overlapping nodes takes $O(n(d+k))$.

For each of our $|C|$ pairs of overlapping nodes we find their common ancestor. The number of steps until we find this is dependent on the height $h$ of the tree. Incrementing counters is no different. Thus, handling overlap takes $O(|C|\cdot h)$. 

Applying the forces is a simple loop over all nodes. Then, updating the window query structure takes $O(d)$ for each updated node. Thus, applying forces and updating nodes takes $O(nd)$.

We get a time complexity of $O(n(d+k) + |C|\cdot h)$ for the three subroutines. Since the number of collisions is $O(n)$ (often much less, especially so in later iterations), we can state that a single iteration has time complexity of $O(n (d + k + h))$.

Finally, since we are handling hierarchical data, it is important to discuss the impact of the branching factor on this analysis. The height of a tree with branching factor $f$ is $O(\log_f(n))$, changing an iteration's time complexity to $O(n (d + k) + n \log_f(n))$. Note that a higher branching factor will likely impact the number of collisions, but not change the bounds of this analysis.

\newpage

\subsection*{Influence of Number of Iterations}

\begin{figure}[h]
    \centering
    \includegraphics[width=\textwidth]{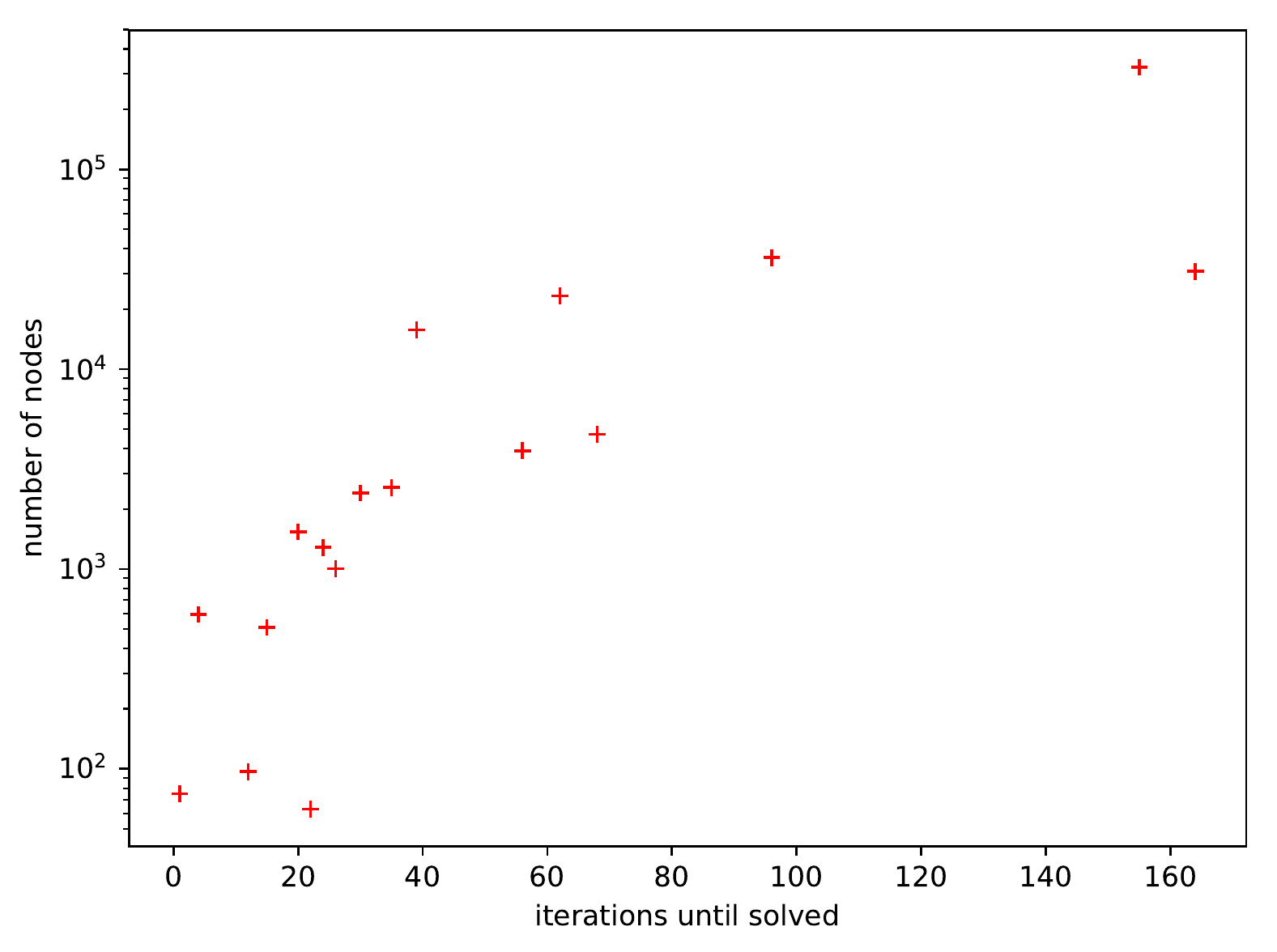}
    \caption{The number of nodes in relation to the number of iterations until stabilization.}
    \label{fig:meta_nodes}
\end{figure}

\begin{figure}[h]
    \centering
    \includegraphics[width=\textwidth]{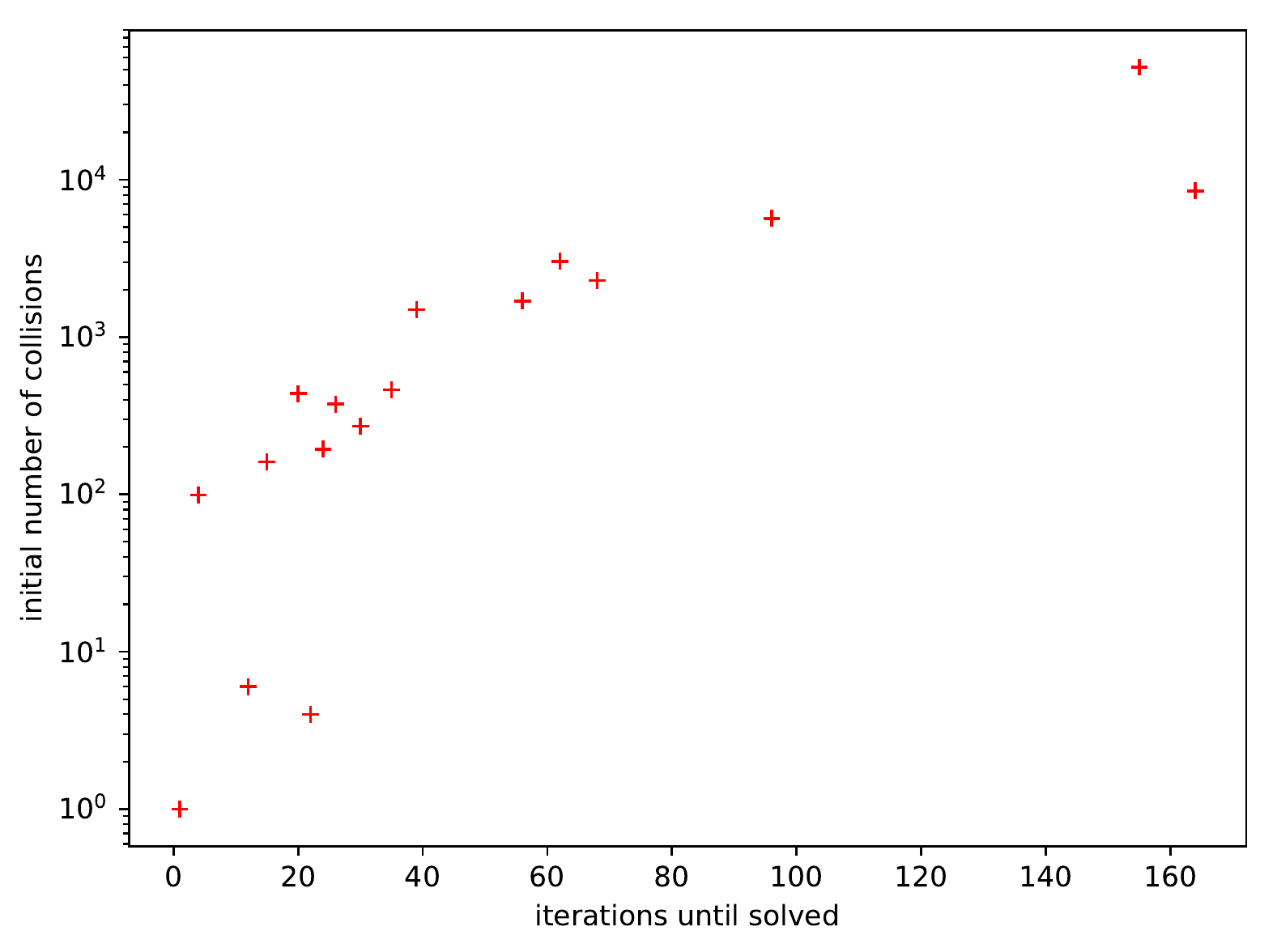}
    \caption{The initial number of collisions in relation to the number of iterations until stabilization.}
    \label{fig:meta_collision}
\end{figure}

\begin{figure}[h]
    \centering
    \includegraphics[width=\textwidth]{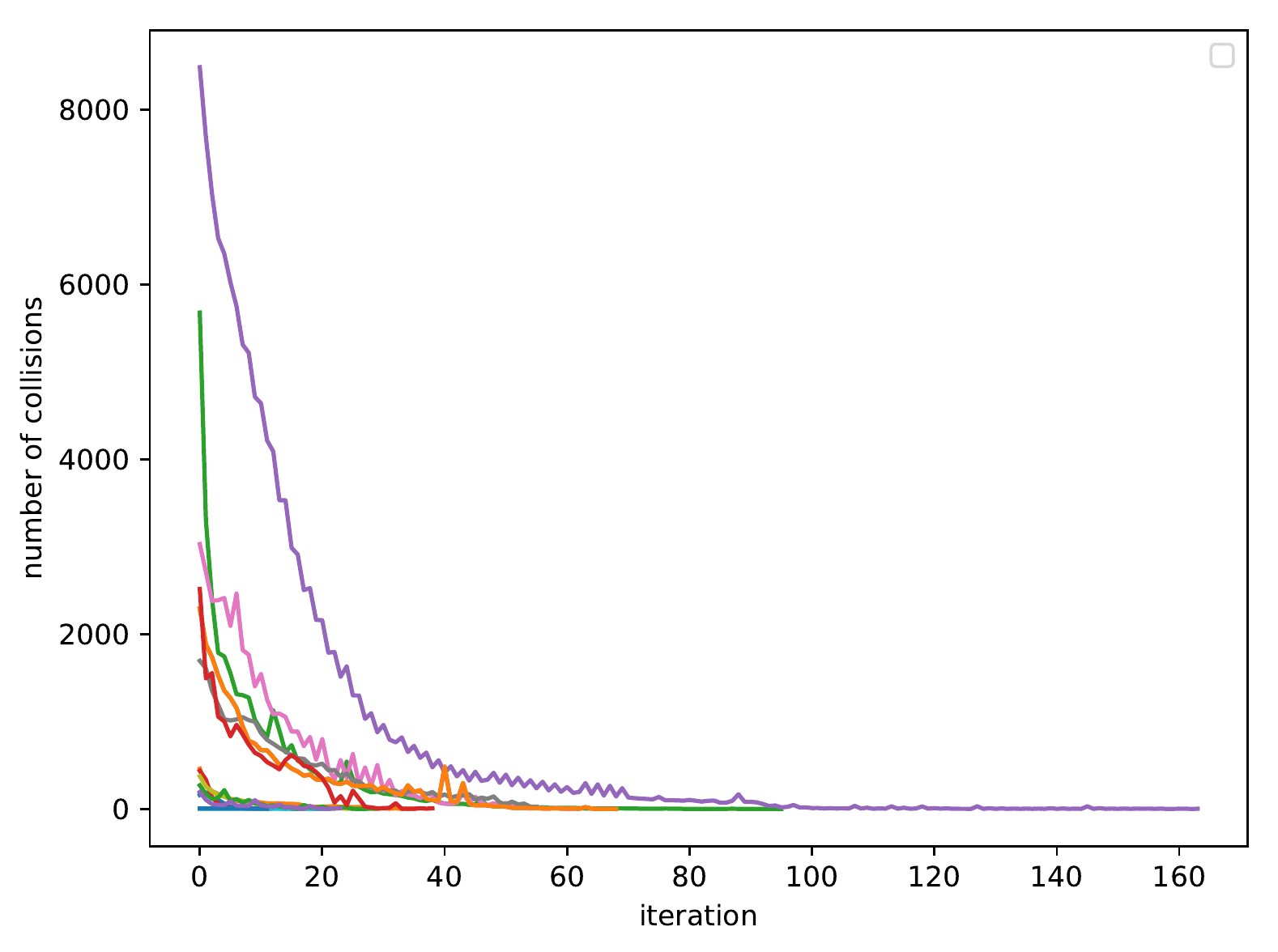}
    \caption{The number of collisions relative to the iteration, for various trees (excluding the NCBI Taxonomy, to keep the $y$-scale manageable).}
    \label{fig:all_relation}
\end{figure}

Figures \ref{fig:meta_nodes}, \ref{fig:meta_collision} and \ref{fig:all_relation} show the relationship between the number of nodes/collisions and the number of iterations until stabilization and the number of collisions relative to each iteration for a selection of trees.

